\documentclass[5p,twocolumn,times]{elsarticle}

\usepackage{lipsum}
\usepackage{lineno,hyperref}
\modulolinenumbers[5]
\usepackage[pdftex]{color}
\usepackage[font=footnotesize,labelfont=bf]{caption}
\usepackage[font=footnotesize,labelfont=bf]{subcaption}
\journal{Journal of \LaTeX\ Templates}

\usepackage{amssymb}

\biboptions{compress}

\usepackage[figuresright]{rotating}

\begin{document}

\begin{frontmatter}
\title{Mobility unevenness in rock-paper-scissors models}

\address[1]{Escola de Ci\^encias e Tecnologia, Universidade Federal do Rio Grande do Norte\\
Caixa Postal 1524, 59072-970, Natal, RN, Brazil}
\address[2]{Institute for Biodiversity and Ecosystem
Dynamics, University of Amsterdam, Science Park 904, 1098 XH
Amsterdam, The Netherlands}

\author[1,2]{J. Menezes}  
\author[1]{S. Rodrigues}  
\author[1]{S. Batista}

\begin{abstract}
We investigate a tritrophic system whose cyclic dominance is modelled by the rock-paper-scissors game. We consider that organisms of one or two species are affected by movement limitations, which unbalances the cyclic spatial game. Performing stochastic simulations, we show that mobility unevenness controls the population dynamics.
In the case of one slow species, the predominant species depends on the level of mobility restriction, with the slow species being preponderant if the mobility limitations are substantial. If two species face mobility limitations, our outcomes show that being higher dispersive does not constitute an advantage in terms of population growth. On the contrary, if organisms move with higher mobility, they expose themselves to enemies more frequently, being more vulnerable to being eliminated.
Finally, our findings show that biodiversity benefits in regions where species are slowed. Biodiversity loss for high mobility organisms, common to cyclic systems, may be avoided with
coexistence probability being higher for robust mobility limitations.
Our results may help biologists understand the dynamics of unbalanced spatial systems where organisms' dispersal is fundamental to biodiversity conservation.
\end{abstract}

%\begin{keyword}
%population dynamics \sep cyclic models \sep stochastic simulations

%\end{keyword}

\end{frontmatter}

%%%%%%%%%%%
\section{Introduction}
\label{sec1}

Mobility plays a vital role in biodiversity maintenance and ecosystem dynamics \cite{ecology,Causes,MovementProfitable,Nature-bio}. Environmental conditions may stimulate organisms' movement, motivating them to search for patches where natural resources indispensable to the perpetuation of the species is available. \cite{ecology,foraging,butterfly,BUCHHOLZ2007401}. Behavioural movement strategies have been observed in many species that respond to environmental changes, adapting their movement following signals received from the neighbourhood \cite{adaptive1,adaptive2,Dispersal,BENHAMOU1989375,coping}. The richness of the animal behavioural has also inspired engineers to create sophisticated tools allowing animats to move strategically in adverse scenarios \cite{animats}.

It has been shown that individuals' mobility may vary according to environmental conditions. For example, the mobility of sulfonamides depends on the pH, and soil charge \cite{sulfo}. Other organisms have their dispersal limited by the topology of the space \cite{diffusionlimited}, e.g., 
plant architecture plays a central role in the success of ladybird beetles in searching for aphids \cite{plantar}. In many cases, different species feel limitations differently, depending on organisms' physical features such as size or the ability to adapt when moving in fragmented landscapes \cite{size,fragmented}.
The difference in mobility between species imposed by environmental constraints
can control population dynamics, determining the chances of species persisting depending on the speed they move \cite{sedentary,urbanisation,dispersion}.

The role of space in biodiversity promotion has also been observed in systems with cyclic dominance among species. Experiments with bacteria \textit{Escherichia coli} showed that biodiversity is maintained if individuals interact locally \cite{Coli,bacteria,Allelopathy}. The spatial interactions among bacteria strains follow the rules of the cyclic, nonhierarchical, rock-paper-scissors game \cite{Coli,bacteria,Allelopathy}.
For this reason, a number of stochastic models have been proposed to study the impact of mobility in cyclic models, both for organisms moving according to the random walk theory or directionally, performing
attack or defence strategies \cite{Reichenbach-N-448-1046,Szolnoki-JRSI-11-0735, Moura, Anti1,anti2,MENEZES2022101606,PhysRevE.97.032415,Avelino-PRE-86-036112}. Furthermore, attention has been given to understanding the effects of the unbalanced selection or reproduction activity among species \cite{uneven,PedroWeak,Weak4}. It has been proved that the coexistence may be jeopardised if the unevenness in the spatial interactions is too strong, revealing the relevance of this issue to biodiversity conservation.

In this work, we investigate a cyclic model whose individuals may be affected by environmental mobility limitations. For this purpose, we propose a version of the spatial rock-paper-scissors model, where organisms of each species may feel the local restrictions differently, which generates dispersal mobility. We aim to describe the spatial patterns, quantifying the interference of the mobility limitations in the typical size of the spatial domains. Our goal is to understand how mobility unevenness influences the risk of organisms being eliminated and compute the advantages of population growth to the species profiting from the unbalanced game. Finally, our objective is to discover if the slowing caused by environmental mobility limitations promotes biodiversity.

The outline of this paper is as follows. The Methods are introduced in 
Sec.~\ref{sec2}, where we describe our models and the implementations of the stochastic simulations. Next, we study the effects of the dispersal unevenness in the spatial partners in 
Sec.~\ref{sec3}. The characteristic length size of the typical spatial domains is quantified in Sec.~\ref{sec4}. In Sec.~\ref{sec5}, the selection risk and species densities are calculated, while coexistence probability is studied in Sec.~\ref{sec6}. Our conclusions and discussion appear in Sec.~\ref{sec7}.

% fig 1 %%%%%%%%%%%%%%%%%%%%%
\begin{figure}
\centering
%%% Please, do not change the scale %%%
\includegraphics[width=45mm]{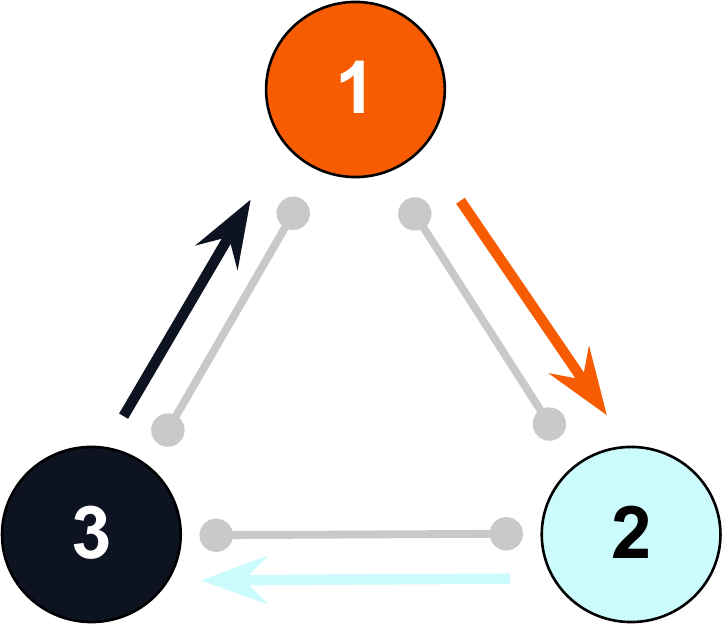}
\caption{Illustration of interaction rules in our cyclic tritrophic system. Red solid lines indicate that organisms of every species compete equally for space. Black, orange, and gray arrows illustrate the cyclic predator-prey interactions.}
	\label{fig1}
\end{figure}
% fig 1 %%%%%%%%%%%%%%%%%%%%%

%%%%%%%%%%%
\begin{figure*}
 \centering
        \begin{subfigure}{.22\textwidth}
        \centering
        \includegraphics[width=40mm]{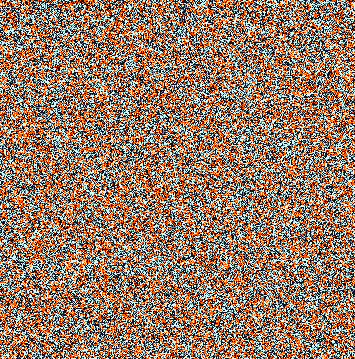}
        \caption{}\label{fig2a}
    \end{subfigure}
       \begin{subfigure}{.22\textwidth}
        \centering
        \includegraphics[width=40mm]{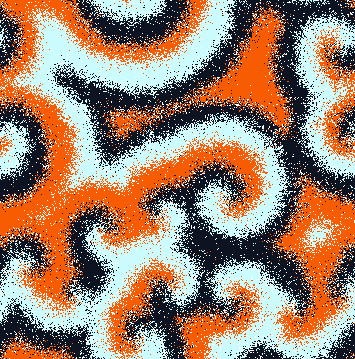}
        \caption{}\label{fig2b}
    \end{subfigure}
   \begin{subfigure}{.22\textwidth}
        \centering
        \includegraphics[width=40mm]{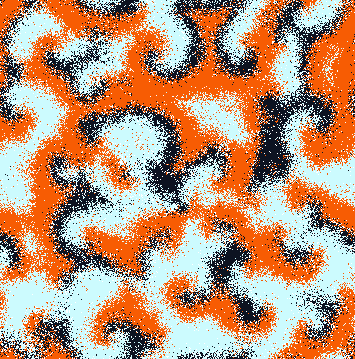}
        \caption{}\label{fig2c}
    \end{subfigure} 
       \begin{subfigure}{.22\textwidth}
        \centering
        \includegraphics[width=40mm]{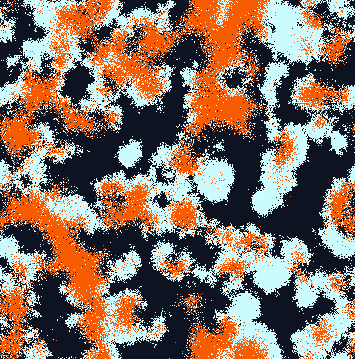}
        \caption{}\label{fig2d}
    \end{subfigure} 
\caption{Snapshots of simulations of the rock-paper-scissors model. The lattices contain $400^2$ grid points, with each organism identified by the colours in Fig.~\ref{fig1} - white dots show the empty spaces.
Figure \ref{fig2a} shows the random initial conditions used in Simulation A, B and C, whose spatial configuration after $3000$ generations are showed in Fig.~\ref{fig2b}, \ref{fig2c}, and \ref{fig2d}, respectively.}
  \label{fig2}
\end{figure*}
%%%%%%%%%%%
%=======================================================================================================

%%%%%%%%%%%
\begin{figure}
 \centering
       \begin{subfigure}{.48\textwidth}
        \centering
        \includegraphics[width=85mm]{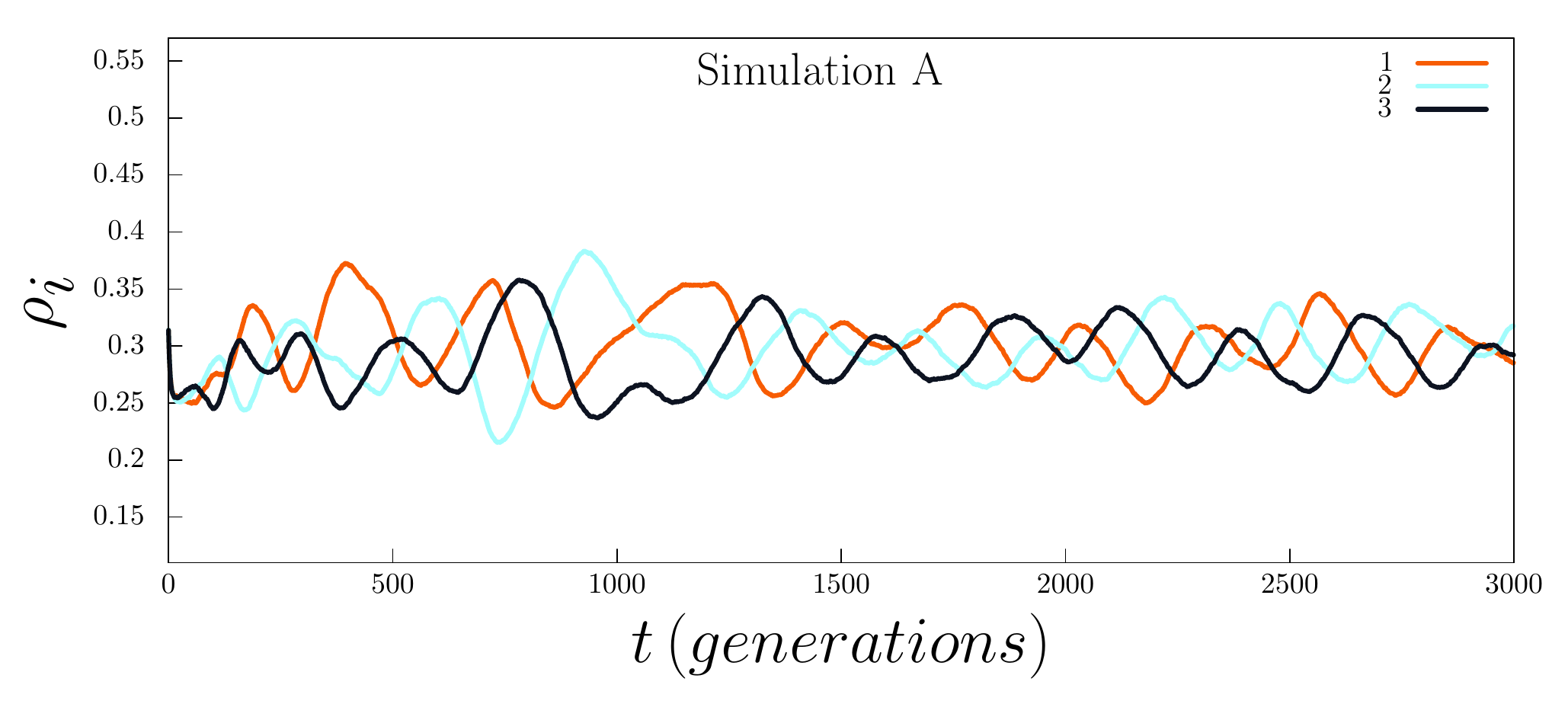}
        \caption{}\label{fig3a}
    \end{subfigure}\\
           \begin{subfigure}{.48\textwidth}
        \centering
        \includegraphics[width=85mm]{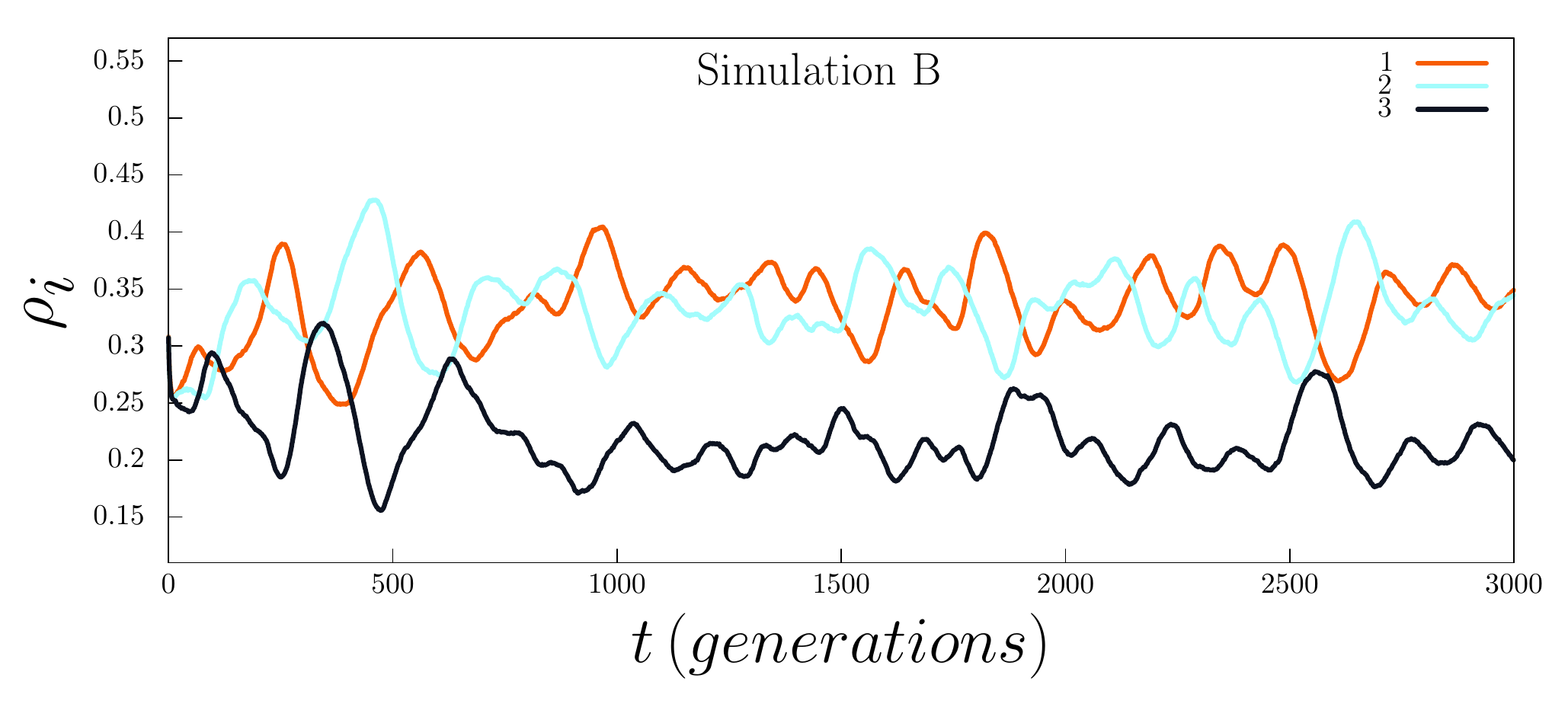}
        \caption{}\label{fig3b}
    \end{subfigure}\\
   \begin{subfigure}{.48\textwidth}
        \centering
        \includegraphics[width=85mm]{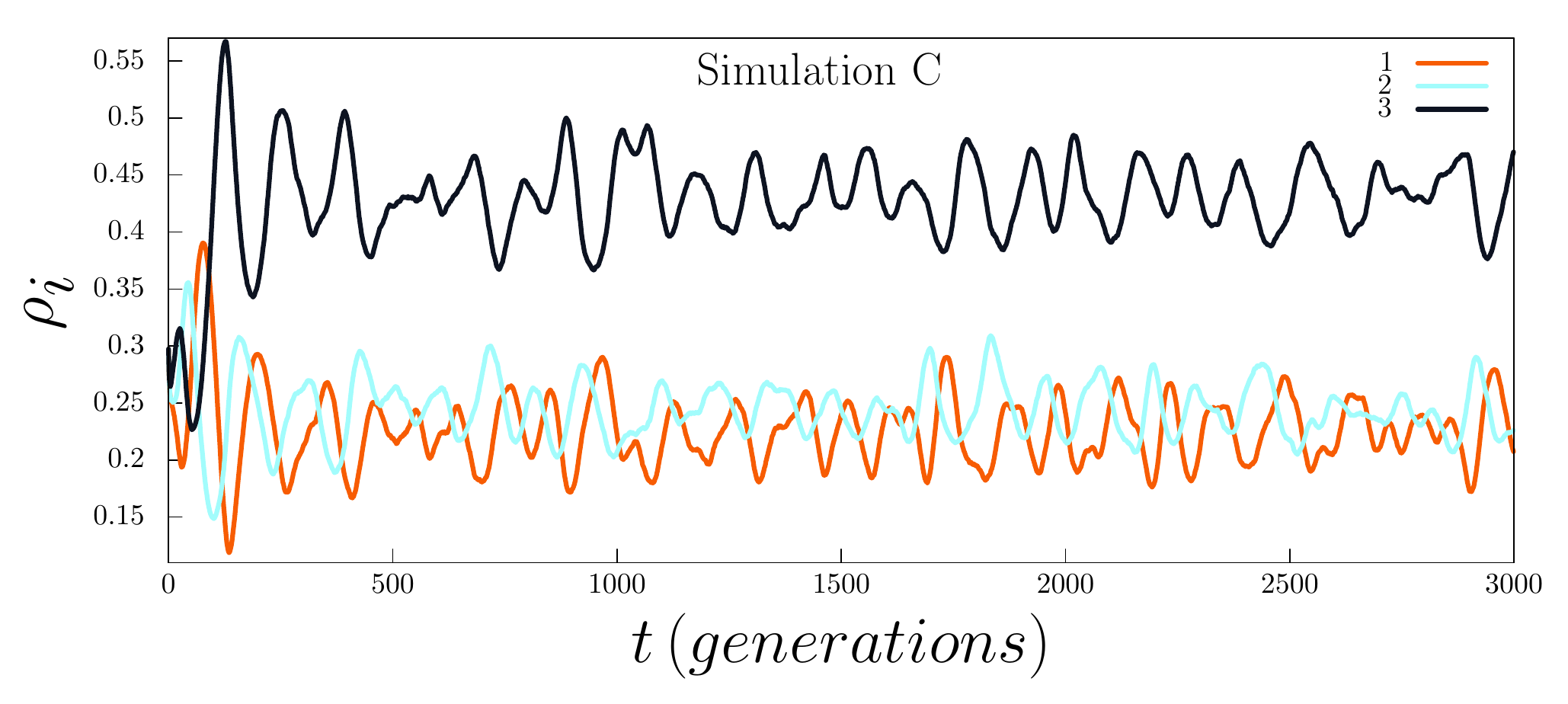}
        \caption{}\label{fig3c}
    \end{subfigure} 
\caption{Dynamics of the species densities. Figures \ref{fig3a}, \ref{fig3b}, and \ref{fig3c} show the fraction of lattices occupied by species $1$ (orange), $2$ (light blue), and $3$ (black) as function of the time for simulations A, B, and C, respectively.}
  \label{fig3}
\end{figure}
%%%%%%%%%%%
\section{Methods}
\label{sec2}

%%%%%%%%%%%
\subsection{The Model}
%%%%%%%%%%%

We study a cyclic tritrophic system whose species dominance is described by the spatial version of the rock-paper-scissors game. Denoting species by $i$, with $i=1,2,3$, we define 
that organisms of species $i$ beat individuals of species $i+1$, with $i=i\,+\,3\,\kappa$ where $\kappa$ is an integer. Figure~\ref{fig1} illustrates the rock-paper-scissors model, where the orange arrow presents the selection interactions occurring if individuals of species $1$ find organisms of species $2$; light blue and black arrows illustrate the dominance of species $2$ and $3$ over species $3$ and $1$, respectively. Light grey bars show the position exchange during a mobility interaction. 
Except for mobility, all interactions occur with the same probability for all organisms of every species.

Mobility probability depends on the intensity of the dispersal limitations suffered by individuals of species $1$ due to the environmental obstruction. 
In this work, we investigate two mobility unevenness cases:
\begin{itemize}
\item
One slow species: mobility probability of organisms of species $1$ is lower, meaning that the changes an organism of species $1$ walking are reduced compared to others.
\item
Two slow species: individuals of species $1$ move with higher probability, i.e., organisms of species $1$ reach further distances than the others in the same time interval. This case is modelled by considering that the species $2$ and $3$ are affected by environmental mobility limitations: species $1$ is referred as the fast species because their organisms are not slowed.
\end{itemize}
%%%%%%%
\subsection{The Simulations}
%%%%%%%

We perform the simulations in square lattices with periodic boundary conditions. Each lattice site contains at most one individual; thus, the maximum number of organisms is $\mathcal{N}$, the total number of grid points. Our numerical implementation follows the May-Leonard model, where the total number of individuals
is not conserved \cite{leonard}. Initially, the total number of organisms is the 
same for every species: $I_i \approx \mathcal{N}/3$, with $i=1,2,3$; each organism is distributed 
at a random grid point. 

At each time step, the spatial configuration is altered by the implementation of one of following interactions:  
\begin{enumerate}
\item \label{ss}
Selection: $ i\ j \to i\ \otimes \,$, with $ j = i+1$, where $\otimes$ means an empty space. Following the rock-paper-scissors
model (Fig.~\ref{fig1}), whenever one selection interaction happens, 
an empty space is left at the position previously occupied by the eliminated organism.
\item \label{rr}
Reproduction: $ i\ \otimes \to i\ i \,$. An individual of any species can use available empty space to reproduce.
\item \label{mm}
Mobility: $ i\ \odot \to \odot\ i\,$, where $\odot$ means an individual of any species. An individual switches positions with another 
individual of any species or goes to an empty space.
\end{enumerate}

Selection and reproduction interactions are implemented with the same
probability for every species; namely, $s$ and $r$, respectively. In contrast, mobility probability depends on the slowness factor $\nu_i$, which represents the dispersal reduction
faced by organisms of species $i$; $\nu_i$ is a real parameter, with $0 \leq \nu_i \leq 1$. In the limit case where mobility of individuals of species $i$ is impracticable, $\nu_i=1.0$; 
if there is no obstruction to move, $\nu_i=0.0$. Therefore, we define the mobility probability for organisms of species $i$ as: $m_i=(1-\nu_i)\,m$, with $m=1-s-r$.
Following this definition,
in our two study cases, the dispersal of individuals of species $i$ is given by: i) Species $1$ is slower: $m_1=(1-\nu)\,m$ and $m_2=m_3=m$; ii) Species $1$ is faster: $m_1=m$ and $m_2=m_3=(1-\nu)\,m$. In both cases, $0 \leq \nu \leq 1$.

Interactions are implemented according to the Moore neighbourhood, where individuals may interact with one of their eight immediate neighbours. The algorithm follows the steps: 
i) randomly choosing an active individual among all organisms in the lattice; 
ii) drawing one interaction to be implemented according to the interaction probabilities; 
iii) raffling one of the eight nearest neighbours to be the passive of the raffled interaction. 
The interaction is executed only if the active and passive fit the conditions
in \ref{ss}, \ref{rr}, and \ref{mm}. In this case, we count one timestep.
Otherwise, we repeat the three steps. When $\mathcal{N}$ interactions
are successfully implemented, one generation is completed (our time unit).
%%%%%
\subsection{Spatial Patterns}
%%%%%
To observe how the spatial patterns are affected by the dispersal unevenness, we ran three simulations starting from the same initial conditions: 
\begin{itemize}
\item
Simulation A: all organisms of every species move without mobility limitations
- the standard model.  We used the set 
of parameters $s=r=0.15$ and $m=0.7$.
\item
Simulation B: only organisms of species $1$ are affected by the environmental dispersal obstructions,
compared to Simulation A; the slowness factor is $\nu=0.1$. 
\item
Simulation C: organisms of species $2$ and $3$ are equally 
slowed with the same slowness factor used in Simulation B.
\end{itemize}
The simulations were performed in lattices with $400^2$ grid sites; the realisations ran for a timespan of $3000$ generations - the temporal dependence of the species densities $\phi_i$ were computed.

%%%%%%
\subsection{Typical Spatial Domain's Characteristic Length}
%%%%%

To compute the scale of spatial domains occupied by each species, we employ the spatial autocorrelation function $C_i(r)$, with $i=1,2,3$, in terms of radial coordinate $r$. For this purpose, we first introduce 
the function $\phi_i(\vec{r})$ that identify the position $\vec{r}$ in the lattice occupied by individuals of species $i$. Using the Fourier transform
\begin{equation}
\varphi_i(\vec{\kappa}) = \mathcal{F}\,\{\phi_i(\vec{r})-\langle\phi_i\rangle\},
\end{equation}
where $\langle\phi_i\rangle$ is the mean value of $\phi_i(\vec{r})$, we find the spectral densities
\begin{equation}
S_i(\vec{k}) = \sum_{k_x, k_y}\,\varphi_i(\vec{\kappa}).
\end{equation}

Next, we calculate the Fourier transform
\begin{equation}
C_i(\vec{r}') = \frac{\mathcal{F}^{-1}\{S_i(\vec{k})\}}{C(0)},
\end{equation}
which can be rewritten as 
a function of the radial coordinate $r$:
\begin{equation}
C_i(r') = \sum_{|\vec{r}'|=x+y} \frac{C_i(\vec{r}')}{min\left[2N-(x+y+1), (x+y+1)\right]}.
\end{equation}
Finally, we define the threshold: $C_i(l_i)=0.15$, where $l_i$ is the characteristic length scale for the spatial domains of species $i$.
%%%%%%%%%%%
\subsection{Selection Risk and Species Densities}
%%%%%%%%%

We investigate how the mobility unevenness influences the chances of an individual of species $i$ being eliminated by organisms of species $i-1$, using the selection risk $\zeta_i$, with $i=1,2,3$. First, we count the number of individuals of species $i$ at the beginning of each generation. Then, we compute how many individuals are killed during the generation. The selection risk is the ratio between the number of eliminated individuals and the initial amount \cite{Moura}.

To quantify the consequences of the organisms' selection risk on the species population, we calculate the spatial densities $\rho$, defined as the fraction of the grid occupied by individuals of the species $i$,
$\rho_{i}(t) = I_i(t)/\mathcal{N}$, where $I_i(t)$ is the total number of individuals of species $i$ at time $t$. The average selection risk $\zeta_i$ and species density $\rho_i$ are found using the outcomes from the second simulation half; this avoids the fluctuations inherent in the pattern formation process.
The selection risk and species density mean value are computed from a set of $100$ simulations, starting from different initial conditions, in lattices with $500^2$ grid points, running for $5000$ generations.

%%%%%%%%%%%%%%%%%%
\subsection{Coexistence Probability}
%%%%%%%%%%%%%%%%%%

To study the impact of mobility unevenness in biodiversity, we compute 
the coexistence probability for a wide range of mobility probability. We
run a set of $1000$ simulations in lattices with $100^2$ grid sites, running until $10000$ generations, for $ 0.05\,<\,m\,<\,0.95$ in intervals of $ \Delta\, m\, =\,0.05$ - selection and reproduction probabilities are given by $s\,=\,r\,=\,(1-m)/2$. 

We investigate the biodiversity loss in the cases of organisms of one or two species being slowed for various values of $\nu$. Coexistence occurs if at least one individual of every species is present when the simulation ends. In other others, coexistence is considered only if $I_i (t=10000) \neq 0$, for $i=1,2,3$. 
Otherwise, our algorithm considers that the simulation results in extinction. Therefore, coexistence probability is the fraction of implementations ending without biodiversity loss.
%%%%%%%%%%%
%%%%%
%%%%%%%%%%%
\begin{figure}
 \centering
       \begin{subfigure}{.48\textwidth}
        \centering
        \includegraphics[width=85mm]{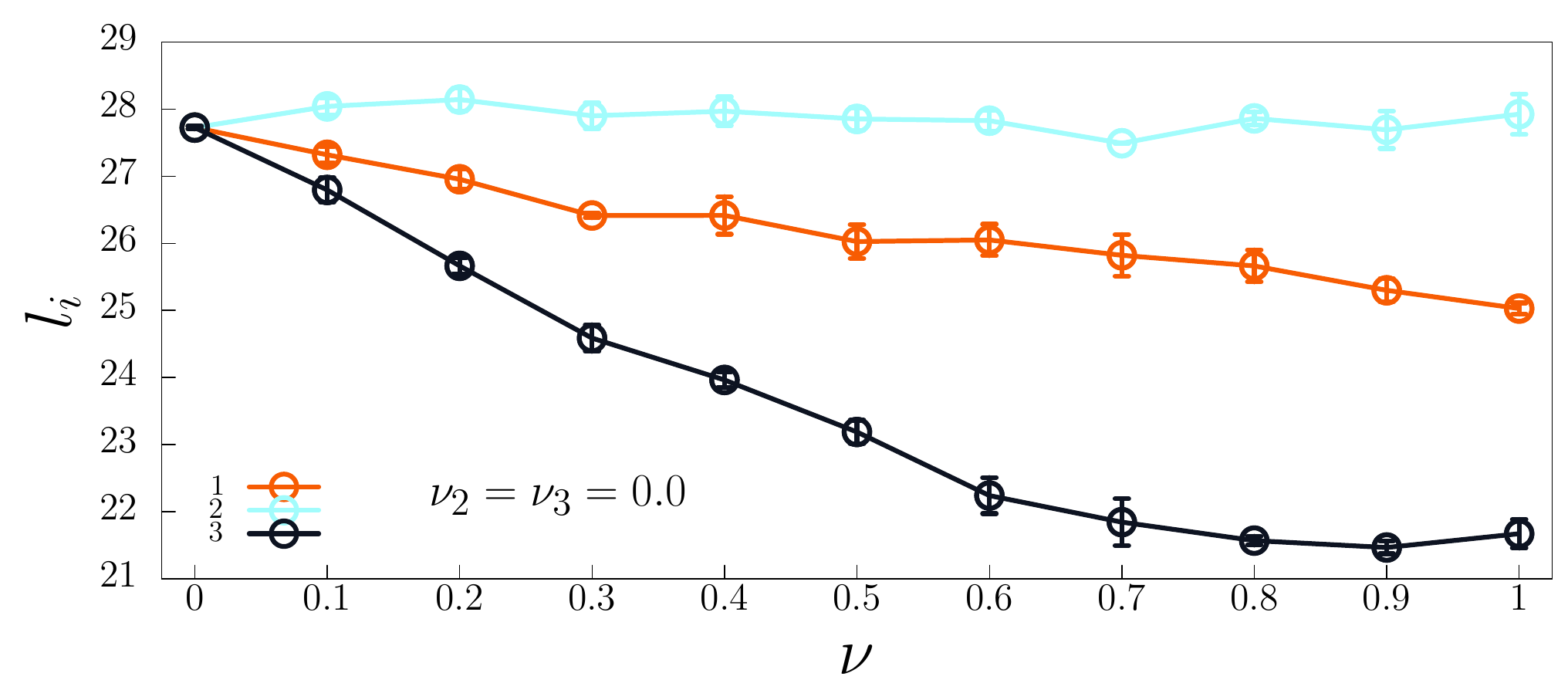}
        \caption{}\label{fig4a}
    \end{subfigure}\\
           \begin{subfigure}{.48\textwidth}
        \centering
        \includegraphics[width=85mm]{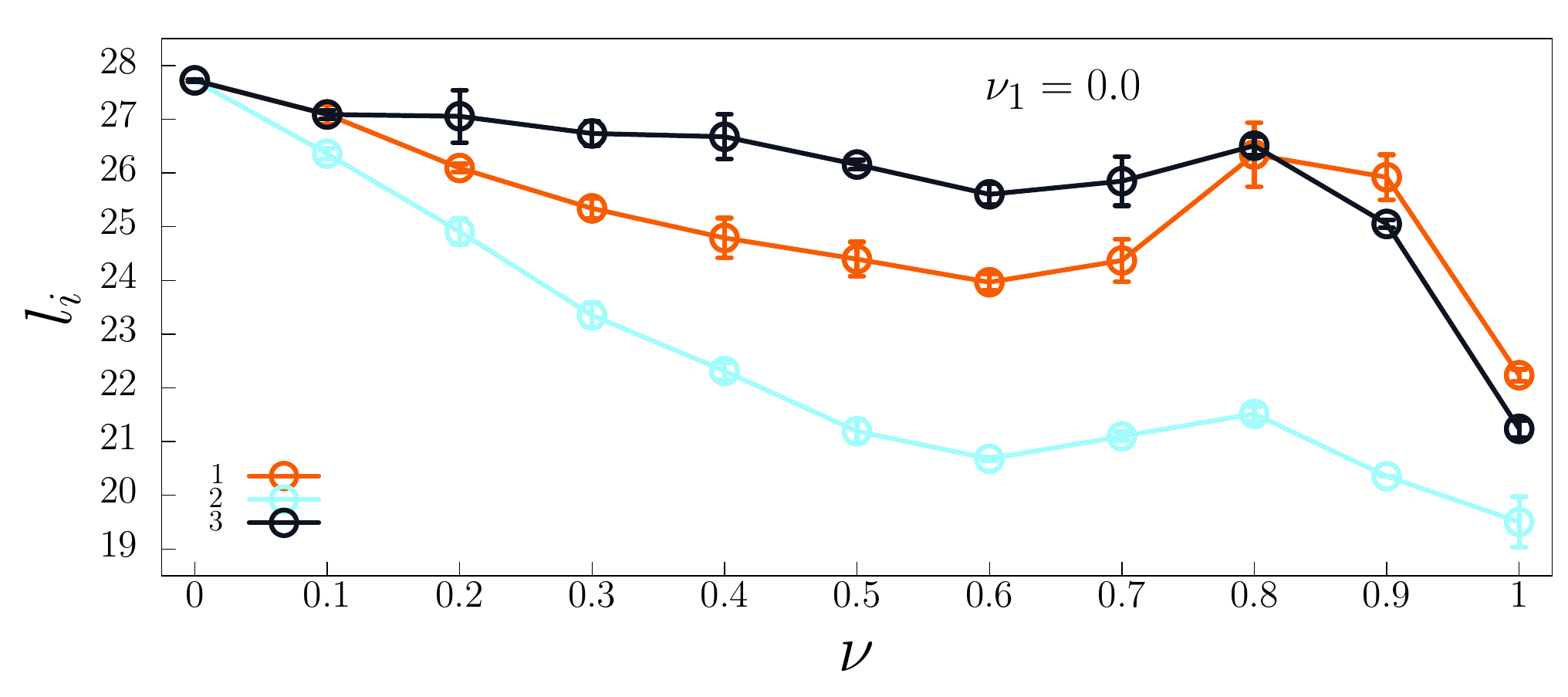}
        \caption{}\label{fig4b}
    \end{subfigure}
\caption{Characteristic length scales of the typical single-species spatial domains. The results were averaged from the sets of $100$ simulations;
the error bars show the standard deviation. Figures ~\ref{fig4a} and \ref{fig4b} depict the outcomes for the case of one slow species and one fast species, respectively. The colours follow the scheme in Fig.~\ref{fig1}.}
  \label{fig4}
\end{figure}
%%%%%%%%%%%
%=======================================================================================================
\section{Spatial Patterns}
\label{sec3}

%================================================================================================================

Let us first study the role of mobility unevenness in the dynamics of spatial patterns. Using the initial conditions shown in Fig. \ref{fig2a}, we ran three simulations whose final spatial configuration appear in the snapshots in Figs. \ref{fig2b} (Simulation A), \ref{fig2c} (Simulation B), and \ref{fig2c} (Simulation C) - see Methods.
The colours follow the scheme in Fig.~\ref{fig1}, with individuals of species $1$, $2$, and $3$ being shown in 
orange, light blue, and black dots, respectively. In addition, empty sites 
are depicted by white dots. Capturing $250$ snapshots of the simulations (in intervals of $20$ generations),
we produced the videos in https://youtu.be/PlTv8pSxOCY (Simulation A), https://youtu.be/zohI-ZZruZY (Simulation B), and https://youtu.be/a9yZCE1ra5A (Simulation C). 
The temporal changes in the species densities $\rho_i$ are depicted in 
Fig.~\ref{fig3a}, \ref{fig3b}, and \ref{fig3c} for simulations A, B, and C, respectively.

After a transient pattern formation stage, regions occupied by individuals of the same species arise. In the standard model, where organisms of every species move without environmental mobility 
limitations, symmetric spiral waves are formed (Fig.~\ref{fig2b}). In this scenario, each species occupies, on average, the same fraction of the grid (Fig.~\ref{fig3a}).
However, there is a remarkable asymmetry in the spatial patterns
in Figs.~\ref{fig2c} and ~\ref{fig2d}.  

To understand the reason for the turbulent pattern formation due to the dispersal unevenness, let us define the flux of organisms per unit time through 
the interface separating the spatial domains of species $i$ and $j$ as $\phi_{i,j}$, with $i,j=1,2,3$. Following the random walk theory \cite{random,Reichenbach-N-448-1046}, we assume that $\phi_{i,j}$ grows with $m_i$ and $m_j$: 
$\phi_{i,j}\,=\,F(m_i, m_j)$, where $F$ is a increasing function in $m_i$ and $m_j$.

\subsection{Simulation B: Slow species}
Because organisms of species $1$ move slower than the others, the $\phi_{2,3} > \phi_{1,2}$  and $\phi_{2,3} > \phi_{1,3}$.
The consequence is that organisms of species $3$ are more vulnerable to being eliminated by individuals of species $2$. This is the reason the black spiral arms are the narrowest (Fig.~\ref{fig2c}), and the average density of species $3$ is the lowest (Fig.~\ref{fig3b}) (see also the video https://youtu.be/zohI-ZZruZY). 

\subsection{Simulation C: Fast species}
In the case of individuals of the species $1$ move faster than the others, the average flux of individuals $\phi_{2,3}$ is much lower than in the other interfaces. Because of this, 
the dynamics of the interface between light blue and black areas are notably slowed; thus, orange regions (the fast species) overgrow, invading the light blue areas (https://youtu.be/a9yZCE1ra5A), being further destroyed by species $2$ that predominates in the cyclic game, as depicted in Figs.~\ref{fig2d} and \ref{fig3c}.

%%%%%%%%%%%
\begin{figure}
 \centering
       \begin{subfigure}{.48\textwidth}
        \centering
        \includegraphics[width=85mm]{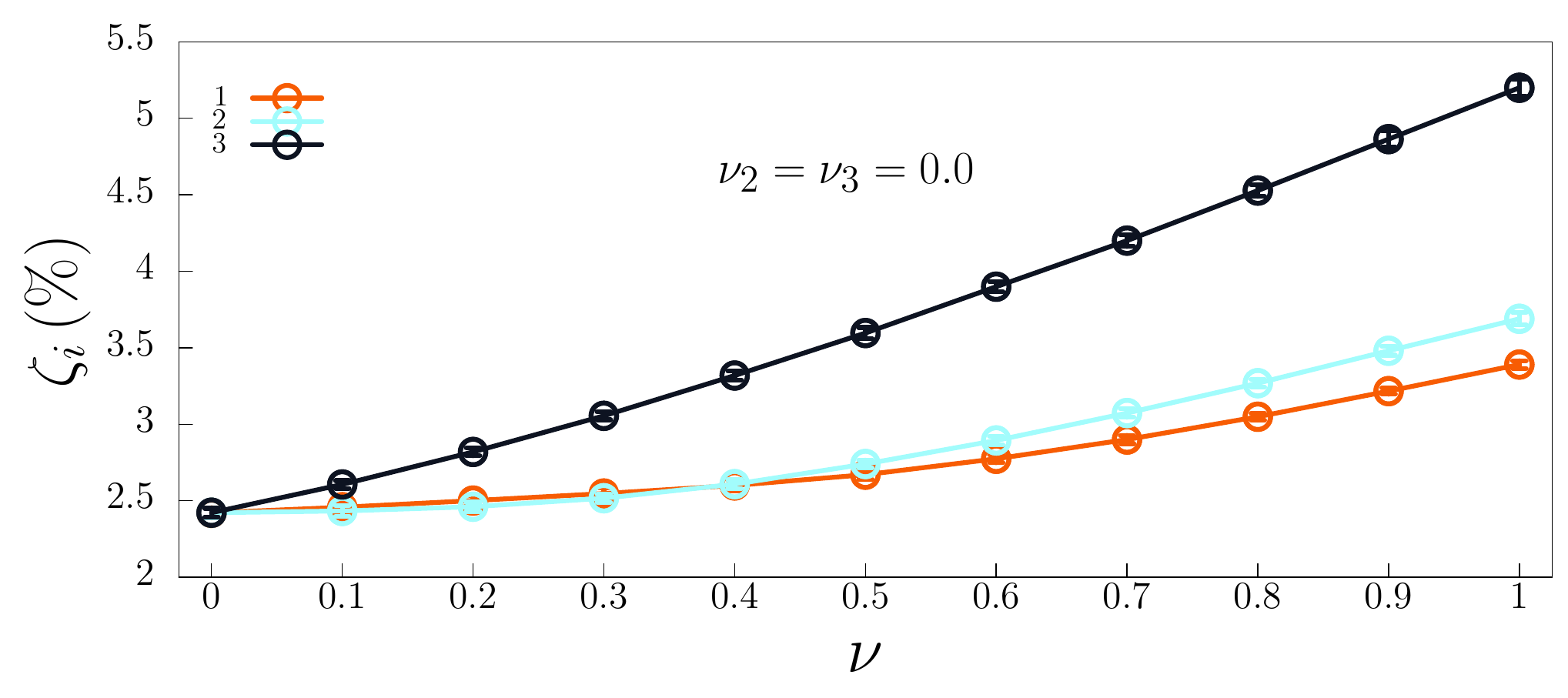}
        \caption{}\label{fig5a}
    \end{subfigure}\\
           \begin{subfigure}{.48\textwidth}
        \centering
        \includegraphics[width=85mm]{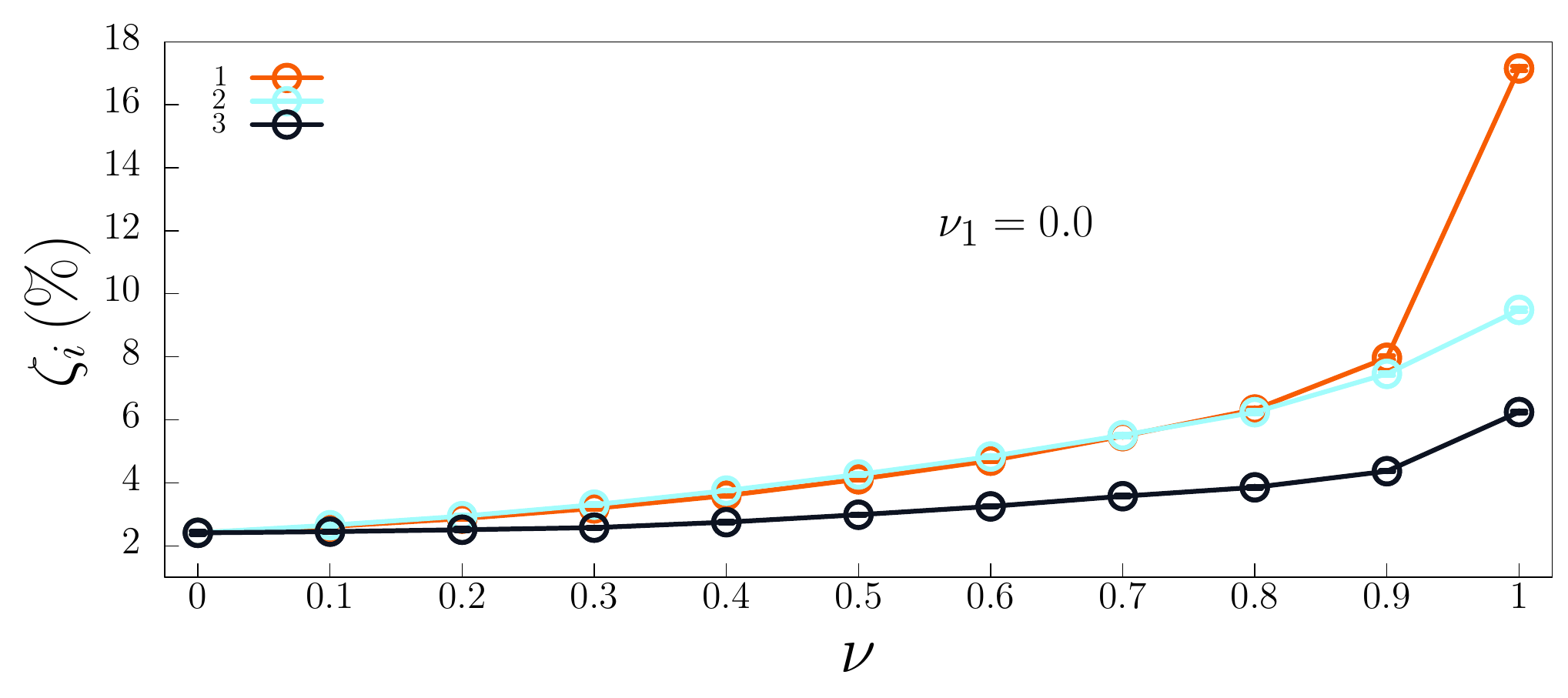}
        \caption{}\label{fig5b}
    \end{subfigure}
\caption{Selection risk as a function of the slowness factor. Figures \ref{fig5a} and \ref{fig5b} show the outcomes for one slow and fast species, respectively. The colours follow the scheme in Fig.~\ref{fig1}; the error bars show the standard deviation.}
  \label{fig5}
\end{figure}
%%%%%%%%%%%

%%%%%%%%%%%
\begin{figure}[t]
 \centering
       \begin{subfigure}{.48\textwidth}
        \centering
        \includegraphics[width=85mm]{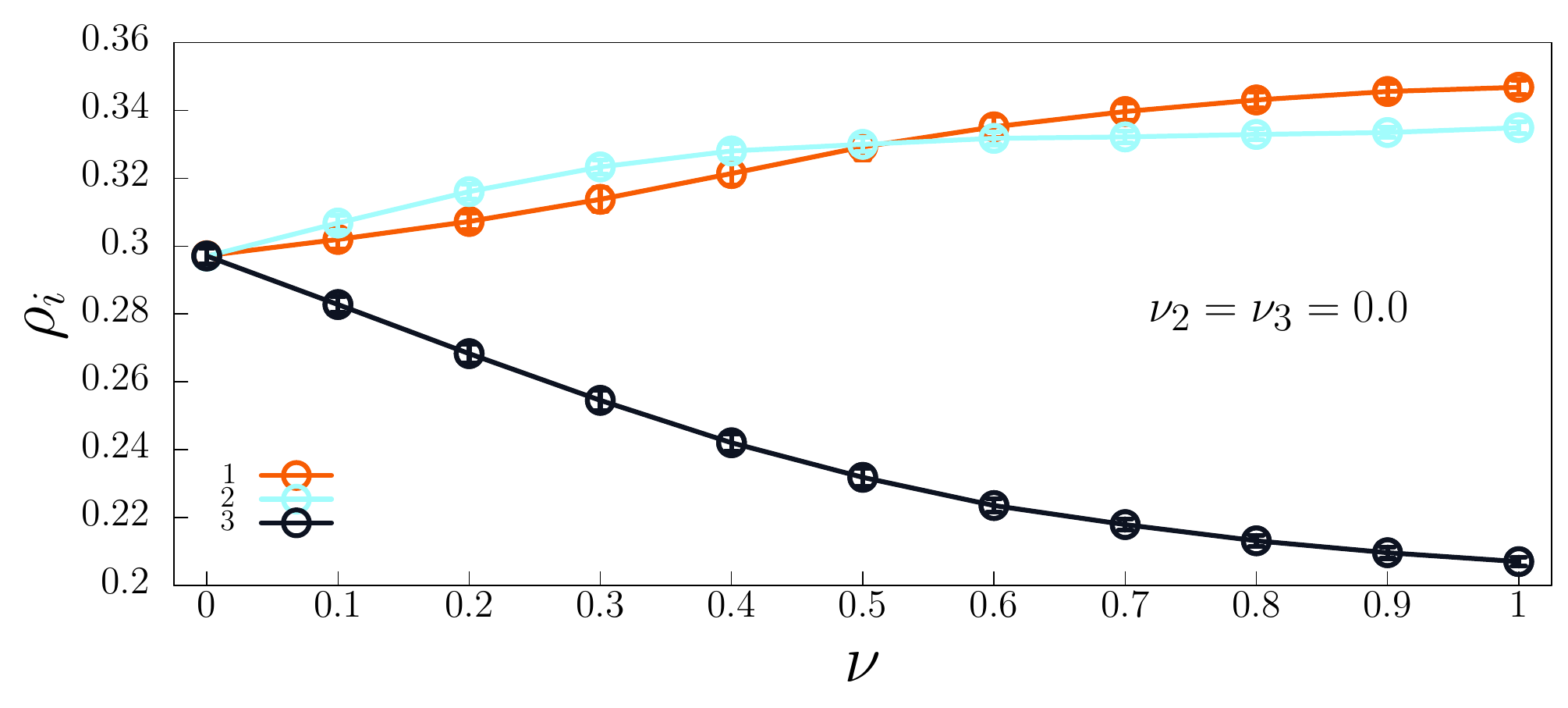}
        \caption{}\label{fig6a}
    \end{subfigure}\\
           \begin{subfigure}{.48\textwidth}
        \centering
        \includegraphics[width=85mm]{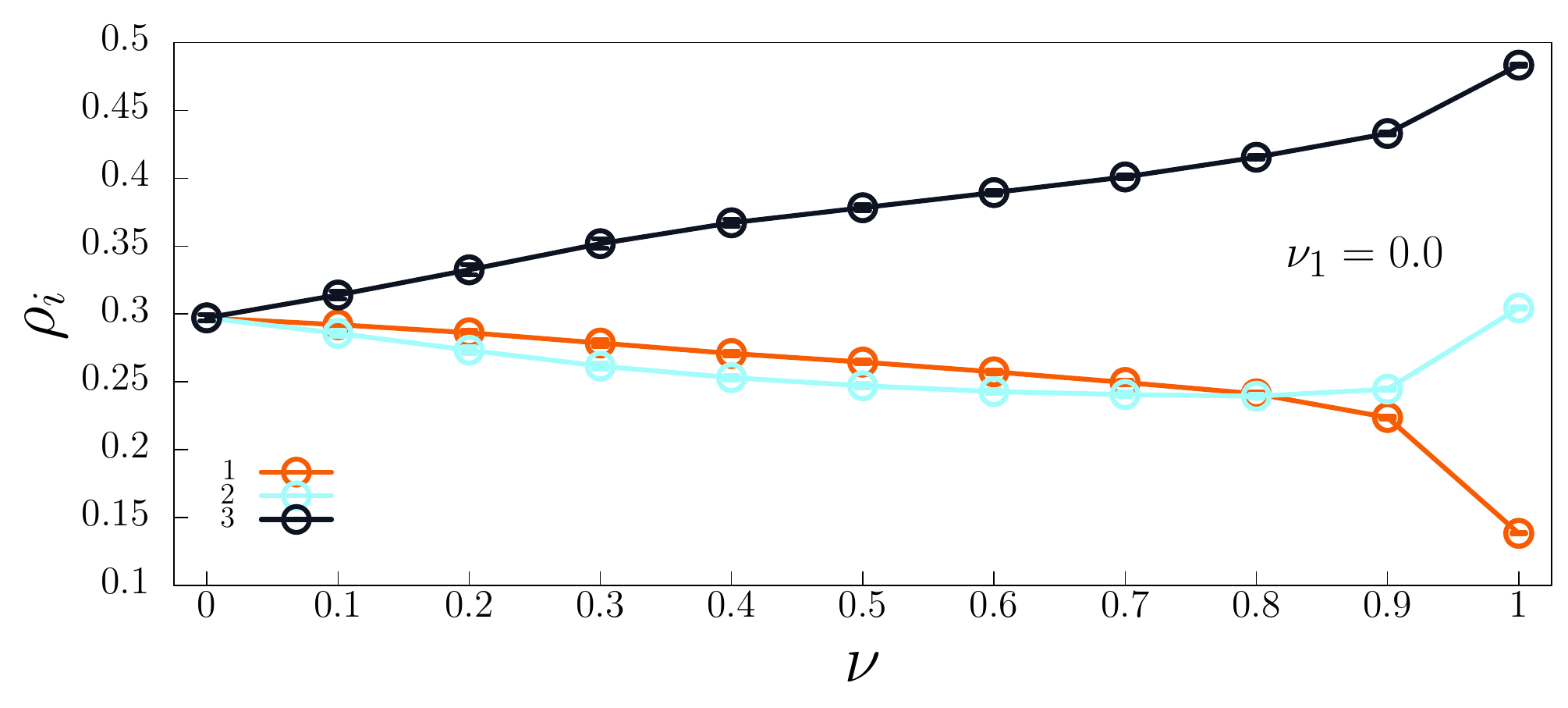}
        \caption{}\label{fig6b}
    \end{subfigure}
\caption{Mean species densities in terms of the slowness factor. Figures \ref{fig6a} and \ref{fig6b} depict the case where individuals of one and two species are slowed, respectively. The colours follow the scheme in Fig.~\ref{fig1}; the error bars show the standard deviation.}
  \label{fig6}
\end{figure}
%%%%%%%%%%%
%=======================================================================================================
\section{Characteristic Length Scale}
\label{sec4}

%=======================================================================================================
%=========

The characteristic length scale of the typical spatial domain occupied by each species was computed by running sets of $100$ simulations. The mean value of $l_i$ appears in Fig.~\ref{fig4a} and \ref{fig4b} for the cases of one and two species with reduced mobility, respectively; the error bars show the standard deviation.

Figure \ref{fig4a} shows that if individuals of species $1$ move slower than the others, the typical sizes of the areas occupied by individuals of species $1$ and $3$ diminish, with $l_3<l_1$, irrespective of $\nu$. Moreover, the lower the dispersal rate of species $1$ is, the shorter $l_1$ and $l_3$ are. On the other hand, the outcomes reveal that the average size of spatial domains of species $2$ is approximately constant with $\nu$. The asymmetry in the spatial patterns reflects the larger number of organisms crossing the interface between spatial domains dominated by species $2$ and $3$. Because of this, 
\begin{itemize}
\item
organisms of species $2$ invade areas dominated by species $3$ faster than individuals of species $3$ manage to conquer the territory of species $1$. Thus, the areas of species $3$ are smaller than the other species;
\item
individuals of species $2$ are more likely to invade areas of species $1$ than being eliminated by species $3$. The result is the increase in $l_2$. 
\item
on both edges of the areas of species $1$, the flux of organisms is approximately the same, but the average number of organisms of species $3$ entering areas of species $1$ is larger than the number of individuals of species $1$ moving to the territory of species $2$. Thus, the typical size of the spatial domains of species $1$ is reduced.
\end{itemize}

In the case of individuals of species $1$ walking with higher mobility, the outcomes depicted in Figure \ref{fig4b} reveal that the typical size of the spatial domains of species $2$ is smaller than the others. This happens because $\phi_{2,3} < \phi_{1,2}$; thus, spatial domains of species $2$ are more invaded than organisms of species $2$ manage to invade areas dominated by species $3$. 
On the other hand, the characteristic lengths of the spatial domains of species $1$ and $3$ depends on the slowness factor: for $\nu \leq 0.8$, $l_1<l_3$; otherwise, $l_1 \geq l_3$.

%=======================================================================================================
\section{Selection Risk and Species Densities}
\label{sec5}

%================================================================================================================
The previous section quantified how the disequilibrium in the flux of organisms on the edges of the single-species domains determines the pattern formation. Now, we aim to understand how mobility unevenness influences the selection risk and the species abundance. Figures \ref{fig5} and \ref{fig6} depict the average value of $\zeta_i$ and $\rho_i$, respectively; the standard deviation is shown by the error bars.

In this case of species $1$ being slow, the flux of organisms is more accentuated in the interface separating areas of species $2$ and $3$; consequently, the risk of being killed is higher for individuals of species $3$ than for other species - as shown in Fig.~\ref{fig5a}. More, as $\nu$ increases, the riskier for individuals of species $3$ is. Therefore, as the dispersal unevenness accentuates, the density of organisms of species $3$ decreases, as depicted by Fig.~\ref{fig6a}.
However, we found that not only organisms of species $3$ face an increased selection risk caused by the mobility reduction of species $1$. Fig.~\ref{fig5a} shows that $\zeta_1$ and $\zeta_2$ also grow when compared with the standard model. Our findings shows a crossover: for $\nu<0.5$, one has $\zeta_1\geq\zeta_2$; otherwise, $\zeta_1<\zeta_2$.
This affects the cyclic game among species, with the species predominance being dependent on $\nu$. Namely, if organisms manage to move less than $50\%$ of the times, species $1$ is preponderant, controlling a higher fraction of the lattice as $\nu$ approaches the maximum value. 

Figure ~\ref{fig5b} shows that, in the case of species $2$ and $3$ with low dispersal, organisms of every species are more susceptible to being selected, with species $1$ and $2$ being more affected than species $3$. This happens
because of the reduced number of organisms crossing the interface 
between territories of species $2$ and $3$.
Due to the lower selection risk, organisms of species $3$ proliferate, as depicted in Fig.~\ref{fig6b}, being the predominant species. Furthermore, although organisms of species $2$ and $3$ face the same mobility limitations, only species $3$ profits with population growth when compared with the standard model.
%=======================================================================================================
\section{Coexistence Probability}
\label{sec6}

%================================================================================================================

\begin{figure}
 \centering
       \begin{subfigure}{.48\textwidth}
        \centering
        \includegraphics[width=85mm]{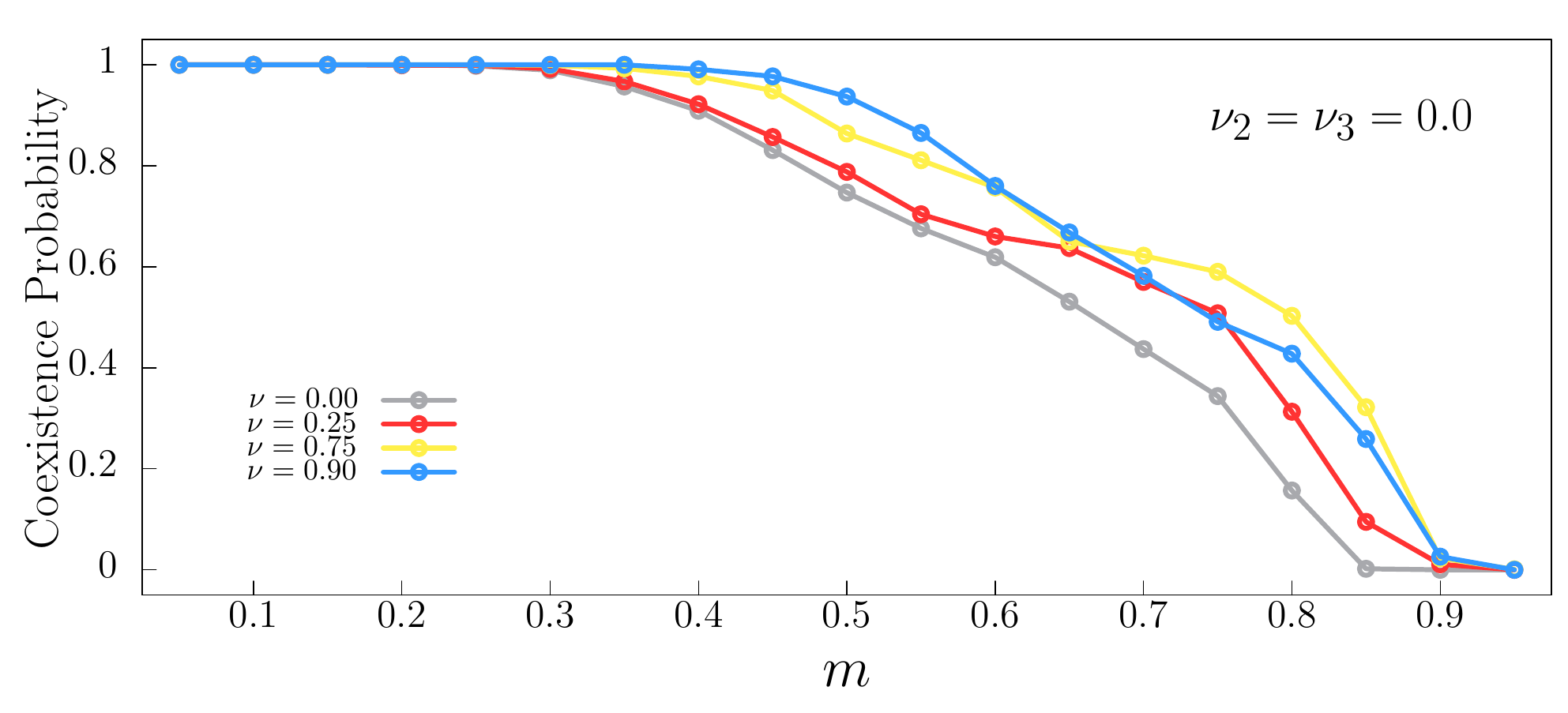}
        \caption{}\label{fig7a}
    \end{subfigure}\\
           \begin{subfigure}{.48\textwidth}
        \centering
        \includegraphics[width=85mm]{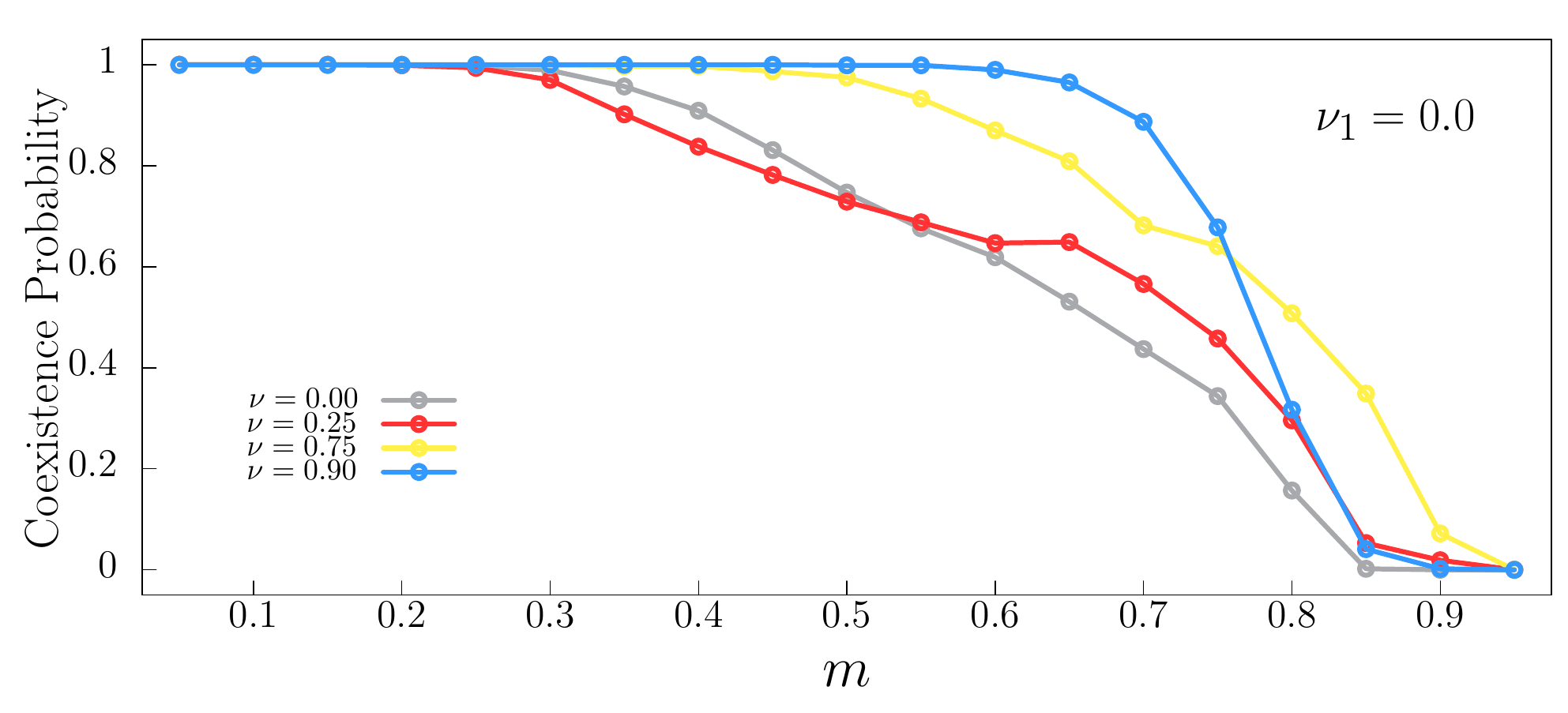}
        \caption{}\label{fig7b}
    \end{subfigure}
\caption{Coexistence probability as a function of the mobility probability $m$. 
Figure \ref{fig7a} and \ref{fig7b} shows the chances of biodiversity holding in the scenarios of organisms of one and two species being affected by mobility limitations for $\nu=0.25$ (red line), $\nu=0.75$ (yellow line), and $\nu=0.9$ (blue line), respectively; grey line depicts the coexistence probability for the standard model.
The results were obtained by running $1000$ simulations in lattices with $100^2$ grid points running until $100^2$ generations.}
  \label{fig7}
\end{figure}

We now investigate the impact of mobility unevenness on biodiversity. Figures \ref{fig7a} and \ref{fig7b} show the coexistence probability in terms of $m$ for the cases of one slow and one fast species - the grey line represents the standard model. Our findings reveal that biodiversity loss in the scenario where individuals move with high mobility is attenuated, with biodiversity benefiting more if more species are affected. Furthermore, the higher the slowness factor is, the higher the chances of species coexisting.

Let us first focus on the case where only individuals of species $1$ are slowed. In this case, biodiversity is benefited
even in the case of weak mobility limitation, as depicted by the red line in Fig. \ref{fig7a} ($\nu=0.25$). Our outcomes show that for $m \leq 0.65$, the coexistence probability increases as the slowness factor grows. This is not valid for $m > 0.65$; in this case, biodiversity is more protected if the slowing provokes a mobility reduction of $75\%$ (yellow line). On the other hand, if organisms of one species move faster than the others, biodiversity is jeopardised for $0.25 \leq m \leq 0.5$ for $\nu \geq 0.25$. The yellow and blue lines show that for $\nu=0.75$ and $\nu=0.9$, the coexistence probability increases, irrespective of $m$. For $m >0.75$, however, biodiversity is more benefited if $\nu=0.75$.

%%%%%%%%%%%
%%%%%%%%%%%

%%%%%%%%%%%%%%%%%%%%%%%%%%%
\section{Discussion and Conclusions}
\label{sec7}
%%%%%%%%%%%%%%%%%%%%%%%%%%%
Our investigation focuses on the spatial version of the rock-paper-scissors game, which is widely used to describe the cyclic interactions among species. We perform stochastic simulations to explore scenarios where organisms of one or two species face environmental mobility restrictions, limiting the dispersal on the lattice.

Because of the mobility unevenness through the interfaces between areas controlled by different species, the symmetric formation of spiral waves is substituted by a turbulent pattern formation process, with species occupying areas with different typical sizes. 
We conclude that if mobility limitations affect only one species, this species benefits since the enemies' selection risk  
increases. This allows the slow species to multiply, being predominant in the case of the species dispersal being reduced to less than $50\%$.
In the scenario of two species being slowed, the fast species' organisms expose themselves more than the others, being caught and eliminated more easily. Our discoveries show that because of the high vulnerability of the organisms of the fast species, the species is the second more abundant if the mobility restrictions impose a mobility reduction not higher than $80\%$; in case of more significant dispersal unevenness, the fast species population occupies the smallest fraction of the grid.

Despite the interference of the mobility unevenness in the population dynamics, being disadvantageous for one or two species (if one or two species are slowed, respectively), biodiversity may be promoted.
Our outcomes reveal that even if organisms commonly disperse with high mobility, the deceleration imposed on the part of the individuals reduces the probability of biodiversity loss. The best results for intermediate mobility are obtained if mobility restrictions are maximum; however, if organisms commonly move with high mobility, the coexistence probability is maximised if the dispersal is diminished to approximately $25\%$. 

In this work, organisms' movement follows the random walk theory \cite{random,Reichenbach-N-448-1046}. However, organisms may move 
towards the most attractive direction if motivated by a behavioural survival strategy \cite{Moura,MENEZES2022101606}. Although some behavioural movement strategies have been studied in the context of cyclic models, environmental mobility restrictions have not been addressed yet. The mobility unevenness may impact the performance of a slow or a fast species when performing gregarious movement, for example, altering the results of the antipredator strategy \cite{MENEZES2022101606}. 

Although we have investigated the scenario where the dispersal of organisms of two species is equally lowered, our main conclusions 
hold for the case where each species is affected differently. The pattern formation process is determined by the number of organisms crossing the interfaces on the borders of the single-species domains. The maximum selection activity happens in the interface where the sum of organisms' mobility probability on both sides is maximum. This determines the least abundant species; following the cyclic selection rules, 
the predominant species can also be determined.
Our results may be helpful for biologists and data scientists to understand species interactions and biodiversity in ecosystems where the same set of species are affected by environmental changes that impose mobility limitations.

%%%%%%%%%%%%%%%%%%%%%%%%%%%%%%%%%%%%%%%%%%%%%%%%%%%%%%%%%%%%%%%%%%%%%%
\section*{Acknowledgments}
We thank CNPq, ECT, Fapern, and IBED for financial and technical support.
%%%%%%%%%%%%%%%%%%%%%%%%%%%%%%%%%%%%%%%%%%%%%%%%%%%%%%%%%%%%%%%%%%%%%%
\bibliographystyle{elsarticle-num}
\bibliography{ref}

\end{document}